\documentclass[aps,prb,twocolumn]{revtex4-2}

\usepackage{graphicx}
\usepackage{upgreek}
\usepackage{amsmath}
\usepackage{amssymb}
\usepackage{dcolumn}
\usepackage{chngpage}

\usepackage{CJK}

\usepackage{graphicx}% Include figure files
\usepackage{dcolumn}% Align table columns on decimal point
\usepackage{bm}% bold math
\usepackage{amssymb} % added by EJF for some math symbols
\usepackage{subfigure}% Include figure files
\usepackage{color}
\usepackage{epstopdf}
\usepackage{tabularx}

\begin{document}

\title{Revisiting the $^{63}$Cu NMR signature of charge order in La$_{1.875}$Ba$_{0.125}$CuO$_{4}$}% Force line breaks with \\

\author{T. Imai$^{1}$}
\author{P. M. Singer$^{2}$}
\author{A. Arsenault$^{1}$}
\author{M. Fujita$^{3}$}
\affiliation{$^{1}$Department of Physics and Astronomy, McMaster University, Hamilton, Ontario, L8S 4M1, Canada}
\affiliation{$^{2}$Department of Chemical and Biomolecular Engineering, Rice University, 6100 Main St., Houston, TX 77005-1892, United States}
\affiliation{$^{3}$Institute for Materials Research, Tohoku University, Sendai 980-8577, Japan}

%\affiliation{%


%\affiliation{%


\date{\today}% It is always \today, today,

\begin{abstract}
We use single crystal $^{63}$Cu NMR techniques to revisit the early $^{63}$Cu NQR signature of charge order observed for La$_{1.875}$Ba$_{0.125}$CuO$_{4}$ ($T_{\text{c}} =4$~K) [A. W. Hunt et al., Phys. Rev. Lett. {\bf 82}, 4300 (1999)].  We show that the growth of spin correlations is accelerated below $\sim 80$~K, where the inverse Laplace transform (ILT)  T$_{1}$ analysis of the $^{139}$La NMR spin-lattice relaxation curve recently uncovered emergence of the slow components in the lattice and/or charge fluctuations [P. M. Singer et al.,  {\bf 101}, 174508 (2020)].   From the accurate measurements of the $^{63}$Cu NMR signal intensity, spin echo decay $M(2\tau)$, spin-lattice relaxation rate $^{63}1/T_1$, and its density distribution function $P(^{63}1/T_{1})$,  we also demonstrate that charge order  at $T_{\text{charge}}\simeq 54$~K turns on strong enhancement of spin fluctuations {\it within charge ordered domains}, thereby making the CuO$_2$ planes extremely inhomogeneous.  The charge ordered domains grow quickly below $T_{\text{charge}}$, and the volume fraction $F_{\text{CA}}$ of the canonical domains unaffected by charge order gradually diminishes by $\sim 35$~K. This finding agrees with our independent estimations of $F_{\text{CA}}$ based entirely on the $^{139}$La ILTT$_{1}$ analyses, but is in a stark contrast with much slower growth of charge ordered domains observed for La$_{1.885}$Sr$_{0.115}$CuO$_{4}$ from its $T_{\text{charge}}\simeq 80$~K to $T_{\text{c}}\simeq 30$~K.     
\end{abstract}

%\begin{abstract}
%An article usually includes an abstract, a concise summary of the work
%covered at length in the main body of the article. 
%\begin{description}
%\end{description}
%\end{abstract}

\maketitle

%\tableofcontents

\section{General Introduction}
A variety of phases compete or coexist in cuprate high $T_c$ superconductors, including the charge ordered phase around the magic composition at $x \sim 1/8$ (see \cite{Robinson2019, FradkinReview2015} for recent reviews).  The charge ordered state was originally discovered a quarter century ago below $T_{\text{charge}} \sim 60$~K in the low temperature tetragonal (LTT) structure of La$_{1.48}$Nd$_{0.4}$Sr$_{0.12}$CuO$_{4}$ \cite{Tranquada1995}.  Years later, evidence for charge order based on neutron and X-ray scattering experiments also emerged in the LTT structure of La$_{1.875}$Ba$_{0.125}$CuO$_{4}$ ($T_{\text{charge}} \simeq 54$~K) \cite{Fujita2004, TranquadaPRB2008, MiaoPRX2019}, followed by La$_{1.68}$Eu$_{0.2}$Sr$_{0.12}$CuO$_{4}$ ($T_{\text{charge}} \sim 80$~K) \cite{Fink}, rather than the low temperature orthorhombic (LTO) structure of the canonical superconducting phase with much higher $T_c$.  Accordingly, many researchers continued to believe that the LTO to LTT structural transformation was the key to stabilizing the long range charge ordered state, which in turn suppresses superconductivity.  
However, recent advances in X-ray scattering techniques finally led to successful detection of charge order Bragg peaks even in the LTO structure of La$_{1.885}$Sr$_{0.115}$CuO$_{4}$ ($T_{\text{c}} \simeq 30$~K) below as high as $T_{\text{charge}} \simeq 80$~K \cite{Croft, Thampy, WenNatComm2019}.  

\begin{figure}
	\begin{center}
		\includegraphics[width=2.8in]{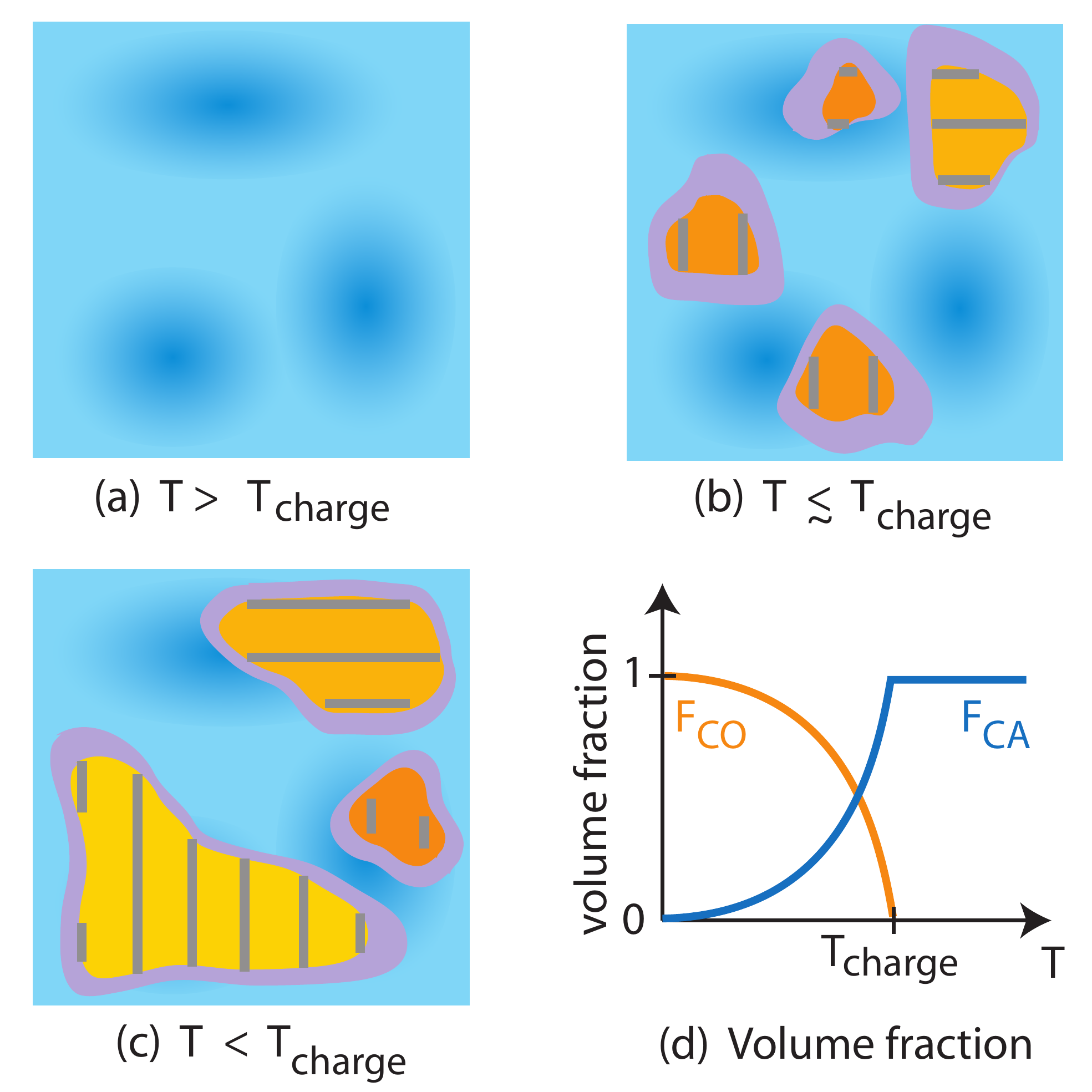}
		\caption{Schematic depictions of two different types of inhomogeneites in the CuO$_{2}$ planes of the La214 type cuprates \cite{HuntPRL1999, SingerPRB1999, HuntPRB2001, SingerPRL2002, SingerPRB2005}.  (a) Above $T_{\text{charge}}$, a mild spatial modulation of the local hole concentration (represented by varying shades) exist with nm length scales, arising from the quenched disorder effects induced by random substitution of Ba$^{2+}$ ions in La$^{3+}$ sites \cite{SingerPRL2002, SingerPRB2005}.   (b) Small charge ordered domains nucleate immediately below $T_{\text{charge}}$ (islands with varying shades, in which the charge rivers are represented by vertical or horizontal lines).  But the charge correlation length is still as short as several nm \cite{TranquadaPRB2008, MiaoPRX2019} and these domains are strongly disordered.  As the size of the charge ordered domain grows (islands with lighter shade), Cu spin fluctuations slow down and hence enhance $1/T_1$.   The disordered nature of the charge ordered state implies that the level of enhancement of $1/T_1$ varies domain by domain, resulting in a large distribution in the enhancement of $1/T_1$.  A majority of Cu sites are not affected by charge order, and maintain the canonical properties expected for superconducting CuO$_2$ planes.  $^{63}$Cu NMR signals in charge ordered domains are observable only with extremely short $\tau \sim 2$~$\mu$s.   On the other hand,$^{63}$Cu NMR signals from the canonical domains are easily observable with $\tau \sim 10$~$\mu$s or longer, but their volume fraction $F_{\text{CA}}$ gradually diminishes below $T_{\text{charge}}$.  (c) With decreasing temperature below $T_{\text{charge}}$, charge ordered domains grow their domain size up to $\sim 20$~nm \cite{TranquadaPRB2008, MiaoPRX2019}, further enhancing $1/T_1$ in larger domains.   (d) The temperature dependence of the volume fractions of the canonical domains $F_{\text{CA}}$ in panels (a-c).  We can also estimate the volume fraction of the charge ordered domains $F_{\text{CO}}$ from $F_{\text{CO}} + F_{\text{CA}} = 1$.
		} 
		\label{fig:domains}
	\end{center}
\end{figure}

Two decades have passed since our initial reports that all of these La214 type cuprates undergo charge order at comparable temperatures \cite{HuntPRL1999, SingerPRB1999, HuntPRB2001, ImaiPRB2018, ImaiJPSJ2018}, on the ground that they all share nearly identical NMR anomalies identified at $T_{\text{charge}}$ of La$_{1.48}$Nd$_{0.4}$Sr$_{0.12}$CuO$_{4}$.  During these years, NMR techniques made major advances both in the instrument technologies and data analysis methods.  Owing to the reduction in the signal detection dead time of the NMR spectrometers after the application of radio frequency pulses, routine NMR measurements have become possible with the pulse separation time as short as $\tau \sim 2$~$\mu$s between the 90 degree excitation and 180 degree refocusing pulses.  This $\tau$ is an order of magnitude shorter than the typical value $\tau \sim 20$~$\mu$s used in the 1980's, and detection of the paramagnetic $^{63}$Cu NMR signals with extremely fast NMR relaxation rates, which arises from the charge ordered domains (represented schematically by islands with various shades in Fig.\ 1(b-c)), has become feasible \cite{PelcPRB2017, ImaiPRB2017}.  Moreover, the development of the inverse Laplace transform (ILT) $T_{1}$ analysis technique enabled us to deduce the histogram of the distribution of the nuclear spin-lattice relaxation rate $1/T_1$ (i.e. the probability density distribution $P(1/T_{1})$) \cite{SingerJCP2018, SingerPRB2019, ArsenaultPRB2019, TakahashiPRX2019}, in addition to  the average value of the distributed $1/T_1$ estimated from the conventional stretched exponential fit.  The recent ILTT$_{1}$ analysis of $^{139}1/T_1$ measured at the $^{139}$La sites in La$_{1.875}$Ba$_{0.125}$CuO$_{4}$ \cite{SingerPRB2019} and La$_{1.885}$Sr$_{0.115}$CuO$_{4}$ \cite{ArsenaultPRB2019} established the continued presence even below $T_{\text{charge}}$ of the canonical domains, which exhibit canonical properties expected for superconducting CuO$_{2}$ planes without anomalous enhancement of Cu spin fluctuations triggered by charge order.  This new finding based entirely on $^{139}$La NMR supports our original conjecture \cite{HuntPRL1999, SingerPRB1999, HuntPRB2001} that peculiar domain-by-domain variation emerges immediately below $T_{\text{charge}}$ due to the spatially growing charge ordered domains, as summarized in Fig.\ref{fig:domains}.

In this paper, we revisit the earlier $^{63}$Cu nuclear quadrupole resonance (NQR) report on the issue of charge order in La$_{1.875}$Ba$_{0.125}$CuO$_{4}$ \cite{HuntPRL1999, HuntPRB2001} based on comprehensive single crystal $^{63}$Cu NMR results, and compare our findings with $^{139}$La NMR results observed for the same crystal \cite{SingerPRB2019}.  Since La$_{1.875}$Ba$_{0.125}$CuO$_{4}$ has a well-defined, sharp charge order transition at $T_{\text{charge}} \simeq 54$~K as determined by X-ray diffraction experiments and lacks magnetic perturbations caused by additional Nd$^{3+}$ spins, it is an ideal platform to test the NMR response that sets in precisely at $T_{\text{charge}}$.  We confirmed a precursor of enhanced spin correlations  below $\sim 80$~K based on the $^{63}$Cu NMR linewidth data \cite{GotoPhysicaB1994} and $1/T_{1}$ \cite{TouLBCOT1}, where the ILTT$_{1}$ analysis of the $^{139}$La NMR data uncovered the presence of low frequency modes in the lattice and/or charge fluctuations \cite{SingerPRB2019}.  These precursors are followed by dramatic, spatially inhomogeneous enhancement of low frequency spin fluctuations {\it within charge ordered domains} that begin to nucleate at $T_{\text{charge}} \simeq 54$~K.  The volume fraction $F_{\text{CO}}$ of the charge ordered domains is {\it not} 100\% immediately below $T_{\text{charge}}$, and grows only progressively below $T_{\text{charge}}$.  We estimate the volume fraction $F_{\text{CA}}={(}1-F_{\text{CO}}{)}$ of the canonical domains based on $^{63}$Cu NMR spin echo decay $M(2\tau)$ measured over a wide time range from $2\tau = 4$ to $100$~$\mu$s.  The temperature dependence of $F_{\text{CA}}$ (Fig.\ref{fig:wipeout}) shows excellent agreement with the independent estimation based entirely on the ILTT$_{1}$ analysis of the $^{63}$Cu NMR $^{63}1/T_1$ data (Fig.\ref{fig:CuILT}) and $^{139}$La NMR $^{139}1/T_1$ data (Fig.\ref{fig:T1ILT})\cite{SingerPRB2019}.  Moreover, we show that these canonical domains almost completely disappear by $\sim 35$~K, far above $T_{\text{c}} \simeq 4$~K of  La$_{1.875}$Ba$_{0.125}$CuO$_{4}$.  This finding is in remarkable contrast with the case of La$_{1.885}$Sr$_{0.115}$CuO$_{4}$, where the canonical domains still occupy nearly a half of the CuO$_2$ planes when superconductivity sets in at $T_{\text{c}} \simeq 30$~K \cite{ImaiPRB2018, ArsenaultPRB2019}. 

The rest of this article is organized as follows.  In section 2, we provide a brief overview of the NMR response expected in the charge ordered CuO$_2$ planes of the La214 cuprates.  In section 3, we present results and discussions, followed by summary and conclusions in section 4.

\section{NMR response in charge ordered La214 cuprates}
Until fairly recently, tremendous confusions persisted since our original publications asserting the presence of charge order in La$_{1.875}$Ba$_{0.125}$CuO$_{4}$ and related La214 materials.  This is primarily because, except for La$_{1.48}$Nd$_{0.4}$Sr$_{0.12}$CuO$_{4}$ \cite{Tranquada1995}, charge order Bragg peaks were not successfully detected in many La214 cuprates for years \cite{Fujita2004, Fink, Croft, Thampy, WenNatComm2019}.  This unfortunate circumstance misled a large number of researchers to argue that charge order was absent in all the cuprates but La$_{1.48}$Nd$_{0.4}$Sr$_{0.12}$CuO$_{4}$, and cast doubt on the link between the $^{63}$Cu NMR anomalies and charge order (see, for example \cite{Curro, JulienPRB2001}).  Moreover, many NMR experts overlooked, or failed to understand the implications of the following crucial statement in our original publication, quoted verbatim from Hunt et al. \cite{HuntPRL1999}:{\it charge order turns on low frequency spin fluctuations \cite{TranquadaPRB59}, and consequently the $^{63}$Cu nuclear spin-lattice and spin-spin relaxation rates diverge in the striped domains}.   

\begin{figure}
	\begin{center}
		\includegraphics[width=2.8in]{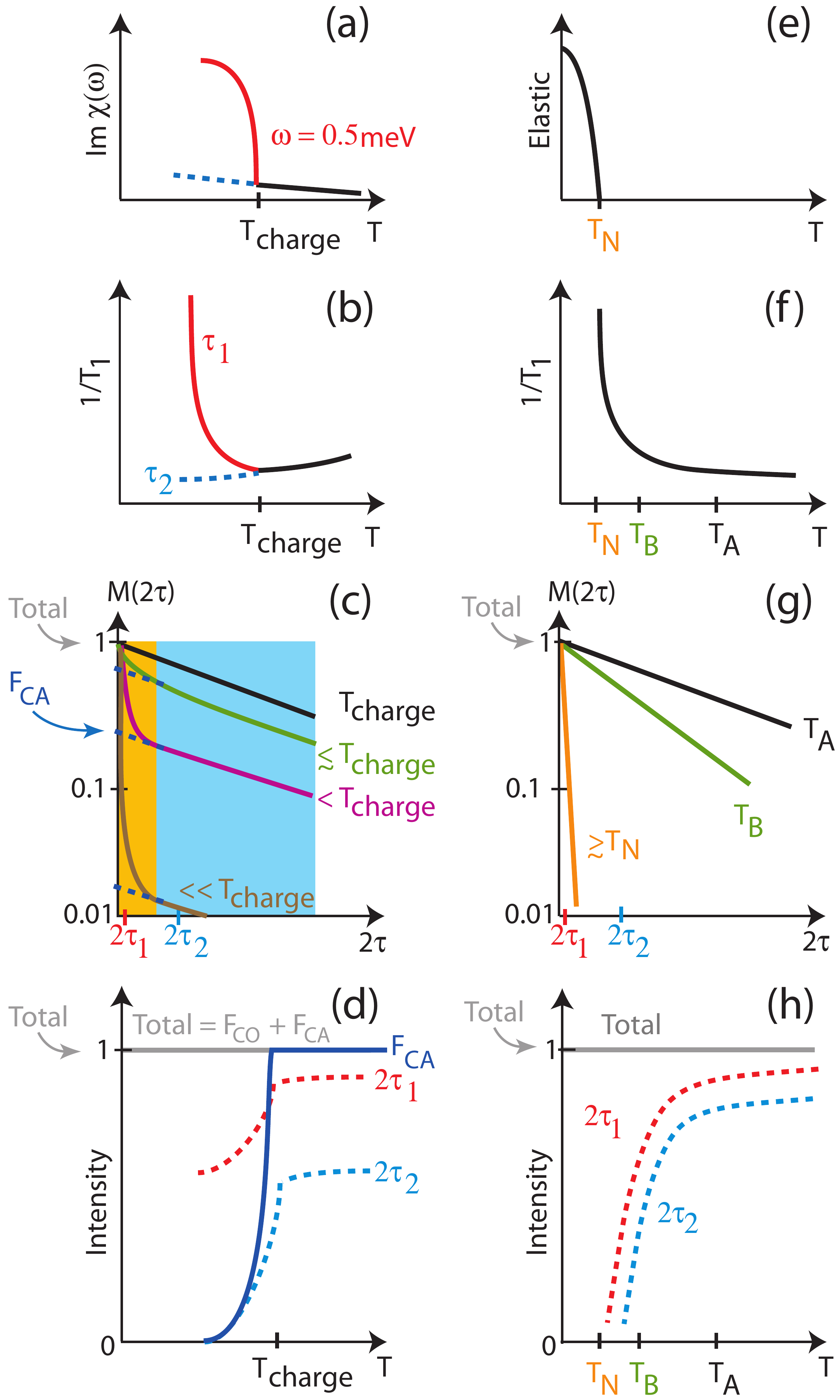}
		\caption{Schematic depictions of the characteristics of charge ordered La214 cuprates (left panels), in comparison to the behavior of conventional, homogeneous antiferromagnets (right panels).   (a) The inelastic neutron scattering intensity at low energy transfers \cite{TranquadaPRB2008}.  See the main text for details.  (b) $^{63}$Cu NMR $1/T_1$ develops a spatial distribution below $T_{\text{charge}}$ with qualitatively different temperature dependences, showing an upturn for short $\tau_1$ but no anomaly for long $\tau_2$.  (c) The Lorentzian (exponential) spin echo decay curve $M(2\tau)$ observed for the $B_{\text{ext}}~||~ab$ axis geometry above $T_{\text{charge}}$ begins to exhibit initial fast decay in the short time regime below $T_{\text{charge}}$, followed by slow decay in the long time regime.   (d) The bare $^{63}$Cu NMR signal intensity measured at a fixed $2\tau_1$ as well as $2\tau_2$ begins to drop at $T_{\text{charge}}$.  The total $^{63}$Cu NMR intensity (gray horizontal line), arising from both the charge ordered and canonical domains, can be estimated by extrapolating $M(2\tau)$ to $2\tau = 0$ in the short time regime in (c).  The total intensity is proportional to the number of $^{63}$Cu nuclear spins in the entire sample, and temperature independent.  On the other hand, the volume fraction $F_{\text{CA}}$ of $^{63}$Cu nuclear spins in the canonical domains, as estimated from the extrapolation of the slowly relaxing component of $M(2\tau)$ in the long time regime of (c), begins to drop below $T_{\text{charge}}$, i.e. $^{63}$Cu NMR signal intensity wipeout effect induced by charge order \cite{HuntPRL1999, HuntPRB2001}.
}
		\label{fig:response}
	\end{center}
\end{figure}

The idea outlined in this short statement about the spatial inhomogeneity of magnetic properties {\it induced by charge order} is the key to the proper understanding of the NMR data of all the La214 cuprates below $T_{\text{charge}}$.  Let us elaborate with the aid of Fig.\ref{fig:response}, in which we contrast the behaviors in the charge ordered state in the left panels with those for typical antiferromagnets without any spatial inhomogeneity, such as La$_2$CuO$_4$ \cite{ImaiPRL1993_1, ImaiPRL1993_2}, in the right panels.  In Fig.\ref{fig:response}(a), we sketch the temperature dependence of the imaginary part of the dynamical local spin susceptibility Im$\chi(\omega)$ at very low energy transfer $\omega$ using solid curves.  See Fig.\ 7 in \cite{TranquadaPRB59} for the first original data for La$_{1.48}$Nd$_{0.4}$Sr$_{0.12}$CuO$_{4}$, and Fig.\ 1(c) in \cite{TranquadaPRB2008} as well as Fig.\ 8(a) in \cite{Fujita2004} for the original data for La$_{1.875}$Ba$_{0.125}$CuO$_{4}$.  In the case of La$_{1.875}$Ba$_{0.125}$CuO$_{4}$, Im$\chi(\omega)$ at the low energy transfer of $\omega = 0.5$~meV begins to grow dramatically precisely at $T_{\text{charge}} \simeq 54$~K \cite{TranquadaPRB2008}.   

This {\it inelastic} glassy spin response induced by charge order below $T_{\text{charge}} \simeq 54$~K should not be confused with the {\it elastic} response of  Bragg scattering arising from the static spin order at $T_{\text{spin}}\simeq 40$~K \cite{Fujita2004, TranquadaPRB2008}.  Simply put, charge order creates small, finite size domains, in which Cu spins are slowly and collectively fluctuating without entering the long range spin ordered state.  The long range spin order takes place when the charge correlation length grows to $\sim 20$~nm at $T_{\text{spin}}\simeq 40$~K \cite{TranquadaPRB2008, MiaoPRX2019}, where charge ordered domains become interconnected.  This sequence of order may look similar to the nucleation of charge density wave (CDW) ordered domains in NbSe$_2$ \cite{STM} above the long range CDW ordered state, but the stripe charge order in La214 materials is accompanied by slow Cu spin fluctuations below $T_{\text{charge}}$.   NMR exhibits spectacular responses to the latter, because NMR is a local, low frequency probe, and hence we can use these NMR responses as the fingerprints of charge order as explained in the following paragraphs.

Since $1/T_{1} \propto T \cdot Im\chi(\omega)/\omega$ at the limit of $\omega \rightarrow 0$\cite{Moriya1963, Jaccarino1965}, one would expect that $1/T_{1}$ shows strong enhancement below $T_{\text{charge}}$, as shown by a solid curve in Fig.\ref{fig:response}(b).  This is somewhat different from the divergent behavior of $1/T_{1}$ toward the N\'eel temperature $T_{N}$ in conventional antiferromagnets shown in panel (f).  In the latter, Im$\chi (\omega)/\omega$ diverges toward $T_{N}$ due to spatially homogeneous critical slowing down of spin fluctuations that begins without any onset temperature, then the elastic response in neutron scattering sets in below $T_{N}$ as shown in Fig.\ref{fig:response}(e).  In the case of charge ordered La$_{1.875}$Ba$_{0.125}$CuO$_{4}$, the elastic magnetic response (Bragg peaks) sets in only below the spin ordering temperature, $T_{\text{spin}} \simeq 40$~K \cite{Fujita2004, TranquadaPRB2008}, but strong enhancement of  $1/T_{1}$ precedes below the clear onset temperature at $T_{\text{charge}}$ due to the glassy nature of spin fluctuations {\it induced by charge order}.

Such an upturn of  $^{139}1/T_{1}$ below $T_{\text{charge}}$ can be easily observed at $^{139}$La sites by NQR \cite{HuntPRB2001}, and more recent high precision NMR measurements confirmed that the onset of the drastic growth of $^{139}1/T_{1}$ is precisely at $T_{\text{charge}} \simeq 54$~K \cite{BaekLaT1PRB2015}.  It is important to note, however, that these earlier $^{139}1/T_{1}$ values were estimated based on the stretched exponential fit, and probed only the average behavior of the entire sample.  In fact, based on the ILTT$_1$ analysis of the $T_1$ recovery curve at $^{139}$La sites, we recently demonstrated that the fastest component of $^{139}1/T_{1}$ indeed begins to grow precisely below $T_{\text{charge}}$ (the solid curve in Fig.\ref{fig:response}(b)), but the slower components exhibit no anomaly through $T_{\text{charge}}$ (dashed curve in Fig.\ref{fig:response}(b)), both in La$_{1.875}$Ba$_{0.125}$CuO$_{4}$ \cite{SingerPRB2019} and La$_{1.885}$Sr$_{0.115}$CuO$_{4}$ \cite{ArsenaultPRB2019}.  

In other words, $1/T_{1}$ exhibits qualitatively different behaviors domain by domain.  Some parts of CuO$_2$ planes are not immediately affected by charge order and continue to exhibit canonical behavior expected for superconducting CuO$_2$ planes even below $T_{\text{charge}}$, as schematically shown in Fig.\ref{fig:domains}(b).  These findings are also consistent with the two decade old knowledge that $1/T_{1}$ measured at $^{63}$Cu sites of La$_{1.875}$Ba$_{0.125}$CuO$_{4}$ with a relatively long $\tau = 20$~$\mu$s exhibits no anomaly at $T_{\text{charge}}$ \cite{ImaiJPSJ1990}, because it preferentially reflects the canonical domains with slow transverse relaxation times, whereas $1/T_{1}$ measured at the $^{63}$Cu sites with very short $\tau$ exhibits an upturn of $1/T_{1}$ below $T_{\text{charge}}$ \cite{TouLBCOT1}. 

The continued presence of $^{139}$La and $^{63}$Cu NMR signals exhibiting the canonical behavior below $T_{\text{charge}}$ implies that not all Cu electron spins are involved in the low energy upturn of Im$\chi(\omega)$ shown by the solid curve in Fig.\ref{fig:response}(a).  Instead, some Cu electron spins continue the trend observed above $T_{\text{charge}}$, as shown by the dashed curve in Fig.\ref{fig:response}(a).  Since inelastic neutron scattering measures only the volume integral of the spin response, one needs to rely on a local probe such as NMR to reveal the domain by domain response schematically summarized in Fig.\ref{fig:domains}.  

The unusual magnetic inhomogeneity induced in the charge ordered state is also reflected on the  transverse $T_2$ relaxation process observed for the transverse nuclear magnetization $M(2\tau)$ at $^{63}$Cu sites, as schematically summarized in Fig.\ref{fig:response}(c) (see Fig.\ref{fig:T2} below for the actual data). Upon entering the charge ordered state, $M(2\tau)$ begins to exhibit initial fast decay in the short time regime, followed by slower decay in the long time regime.  The transverse relaxation rate in the latter is comparable to that observed at $T_{\text{charge}}$ and above.  The nuclear spins responsible for the fast and slow transverse relaxation in Fig.\ref{fig:response}(c) can be attributed to the $^{63}$Cu sites located in the charge ordered and canonical domains in Fig.\ref{fig:domains}, respectively.  

We can estimate the volume fraction $F_{\text{CA}}$ of the canonical domains by extrapolating the slow decaying part of the $M(2\tau)$ curve to $2\tau = 0$, as shown by dashed lines in Fig.\ref{fig:response}(c).  The intercept of the extrapolated dashed line with the vertical axis at $2\tau =0$ yields $F_{\text{CA}}$.  We emphasize that, if the charge ordered CuO$_2$ planes undergo uniform enhancement of spin correlations, then $M(2\tau)$ curve would look very different, and should be similar to the case of uniform antiferromagnets shown in Fig.\ref{fig:response}(g).  

In Fig.\ref{fig:response}(d), we summarize the temperature dependence of the $^{63}$Cu NMR signal intensity at short $2\tau_{1}$ and long $2\tau_{2}$,  as expected from Fig.\ref{fig:response}(c).  The {\it total}  intensity in the limit of $2\tau = 0$ is proportional to the number of nuclear spins that does not change with temperature, and hence always conserved, as shown by the horizontal gray line.  For the finite values of $2\tau$, the intensity exhibits an anomaly at $T_{\text{charge}}$, because the signal intensity arising from the charge ordered domains is reduced by the fast transverse relaxation in the short time regime in Fig.\ref{fig:response}(c).  As explained in the previous paragraph, one can estimate the volume fraction of the canonical domains $F_{\text{CA}}$ by extrapolating $M(2\tau)$ in the long time regime of Fig.\ref{fig:response}(c) to $2\tau=0$.  The end result would be the solid curve in Fig.\ref{fig:response}(d).  Recalling $F_{\text{CO}}+F_{\text{CA}}=1$, one can also estimate $F_{\text{CO}}$ as shown in Fig.\ref{fig:domains}(d).  This is the technique of the signal intensity wipeout effect to probe the volume fraction of the charge ordered domains, developed originally in \cite{HuntPRL1999, SingerPRB1999, HuntPRB2001}.

\section{Results and Discussions}

\subsection{NMR lineshapes}

\begin{figure}
	\begin{center}
		\includegraphics[width=3in]{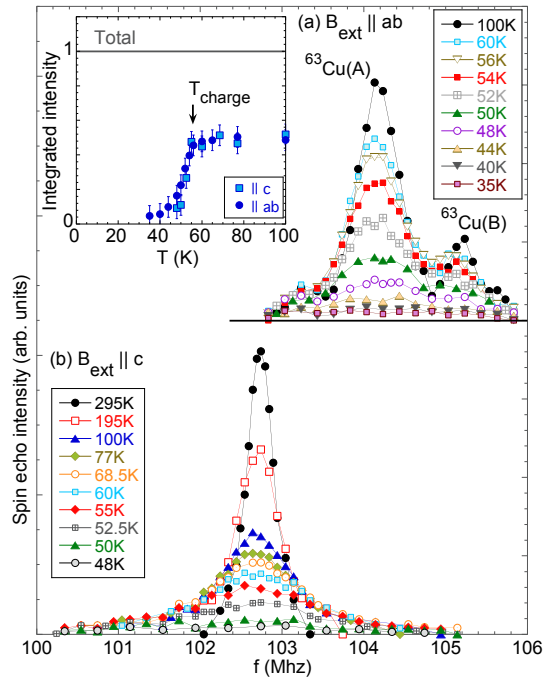}
		\caption{(a) $^{63}$Cu NMR lineshapes observed with $\tau = 12$~$\mu$s for the nuclear spin $I_{\text{z}}=+1/2$ to $-1/2$ central transition in $B_{\text{ext}}=9$~T applied along the (a) ab-plane and (b) c-axis.  The signal intensity is normalized for the Boltzmann factor by multiplying temperature $T$.  The peak frequency in (b) is dominated by the temperature independent chemical shift $\sim 1.2$~\% \cite{PenningtonPRB1989tensor}.  The double peak structure in (a) arises from the difference in the nuclear quadrupole frequency $\nu_{\text{Q}}$ \cite{YoshimuraJPSJ1989, SingerPRL2002}.  The $^{63}$Cu(B) sites have a larger $\nu_{\text{Q}}$ due to the presence of a Ba$^{2+}$ ion at the nearest neighbor La$^{3+}$ sites than the majority $^{63}$Cu(A) sites.  (Inset) The temperature dependence of the integrated intensity of the lineshapes in panels (a) and (b).  Note that these are just raw data without any analysis, and correspond to the dashed curve in Fig.\ref{fig:response}(d) with $2\tau = 24$~$\mu$s.  We estimated the total intensity in the limit of $2\tau = 0$ (defined as 1 after normalization) from the spin echo decay curves above $T_{\text{charge}}$. }
		\label{fig:lineshape}
	\end{center}
\end{figure}

In Fig.\ref{fig:lineshape}(a-b), we summarize the representative $^{63}$Cu NMR lineshapes for a 51~mg single crystal \cite{Fujita2004} measured with a fixed pulse separation time $\tau = 12$~$\mu$s in an external magnetic field $B_{\text{ext}}=9$~T.  The typical radio frequency pulse width was 2~$\mu$s and 4~$\mu$s for 90 and 180 degree pulses throughout this work.  We confirmed both above (60~K) and below (50~K) $T_{\text{charge}} \simeq 54$~K that the lineshape hardly changes even if we use $\tau =2$~$\mu s$, except that the transverse relaxation process (i.e. $T_2$) reduces the overall intensity for longer $\tau$.  These lineshapes indicate that La$_{1.875}$Ba$_{0.125}$CuO$_{4}$ develops its charge ordered state in a fundamentally different manner from La$_{1.885}$Sr$_{0.115}$CuO$_{4}$.  We recall that the c-axis $^{63}$Cu NMR lineshapes in the latter comprised of two distinct types of signals below its higher $T_{\text{charge}}\simeq 80$~K: (i) a narrower, canonically behaving peak with slower relaxation rates that are typical for high $T_c$ cuprates, and (ii) a much broader wing-like signal with extremely fast relaxation rates \cite{ImaiPRB2018}.  The former is gradually wiped out below $T_{\text{charge}}$, transferring the spectral weight to the latter.  Accordingly, the NMR lineshapes completely change between $\tau =2$~$\mu s$ and $\tau =12$~$\mu s$ below $T_{\text{charge}}$, since only the canonically behaving narrower peak can be detected with $\tau =12$~$\mu s$.  That is not the case here for La$_{1.875}$Ba$_{0.125}$CuO$_{4}$, and the observed c-axis lineshapes are more uniformly broadened even for $\tau =12$~$\mu s$. 

In the inset of Fig.\ref{fig:lineshape}, we also summarize the temperature dependence of the integral of these $\tau = 12$~$\mu$s lineshapes.  We emphasize that the intensity data are merely the integral of the lineshapes in panels (a) and (b), and have not been subjected to any data analysis.  The sharp anomaly observed at $\sim 54$~K in the raw integrated intensity  corresponds to that in the conceptual sketch of the dashed curve in Fig.\ref{fig:response}(d) for $2\tau_{2} = 24$~$\mu$s, and signals a lurking phase transition in La$_{1.875}$Ba$_{0.125}$CuO$_{4}$.  

\subsection{Linewidth}

\begin{figure}
	\begin{center}
		\includegraphics[width=3.4in]{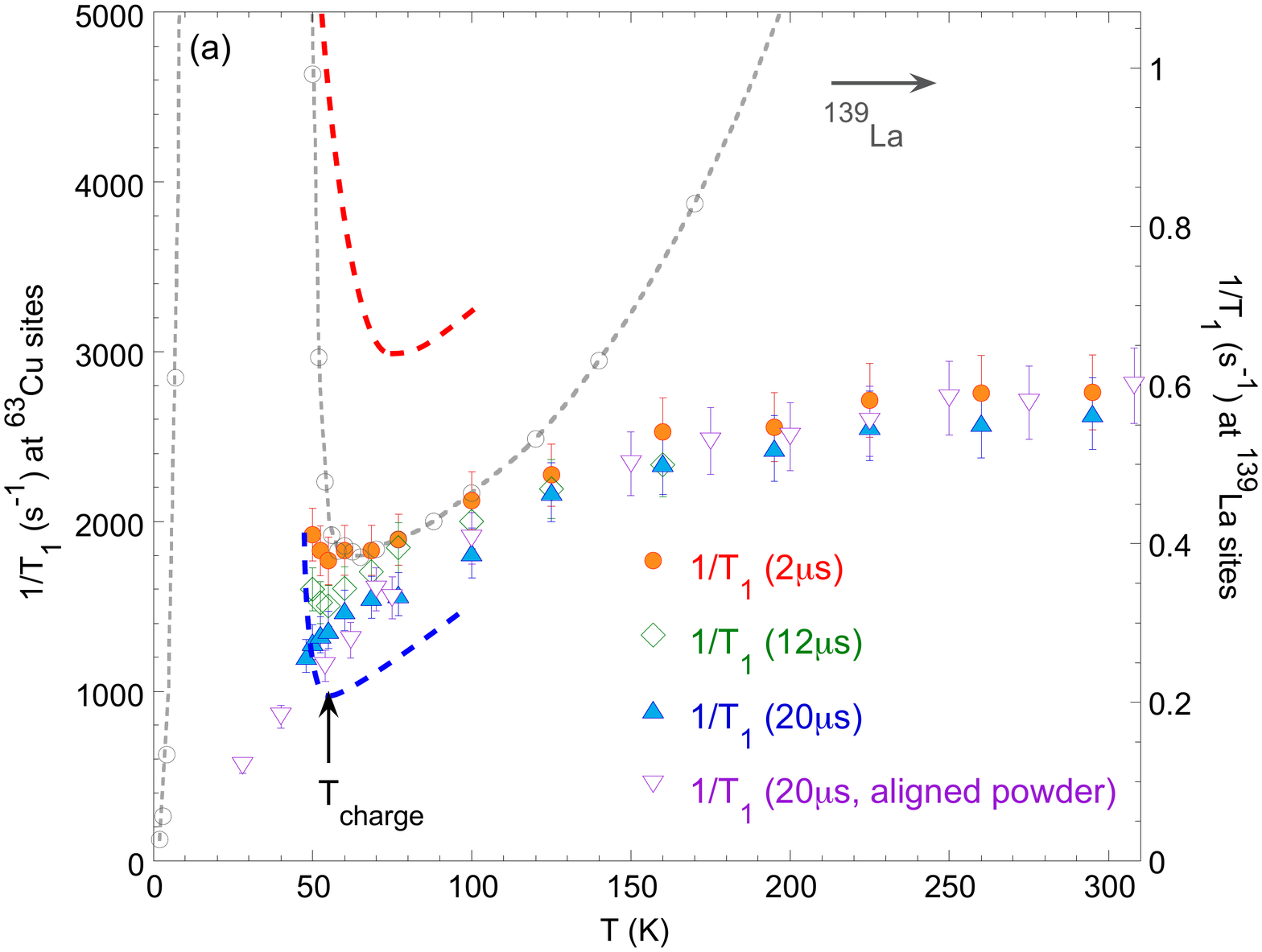}
		\includegraphics[width=3.2in]{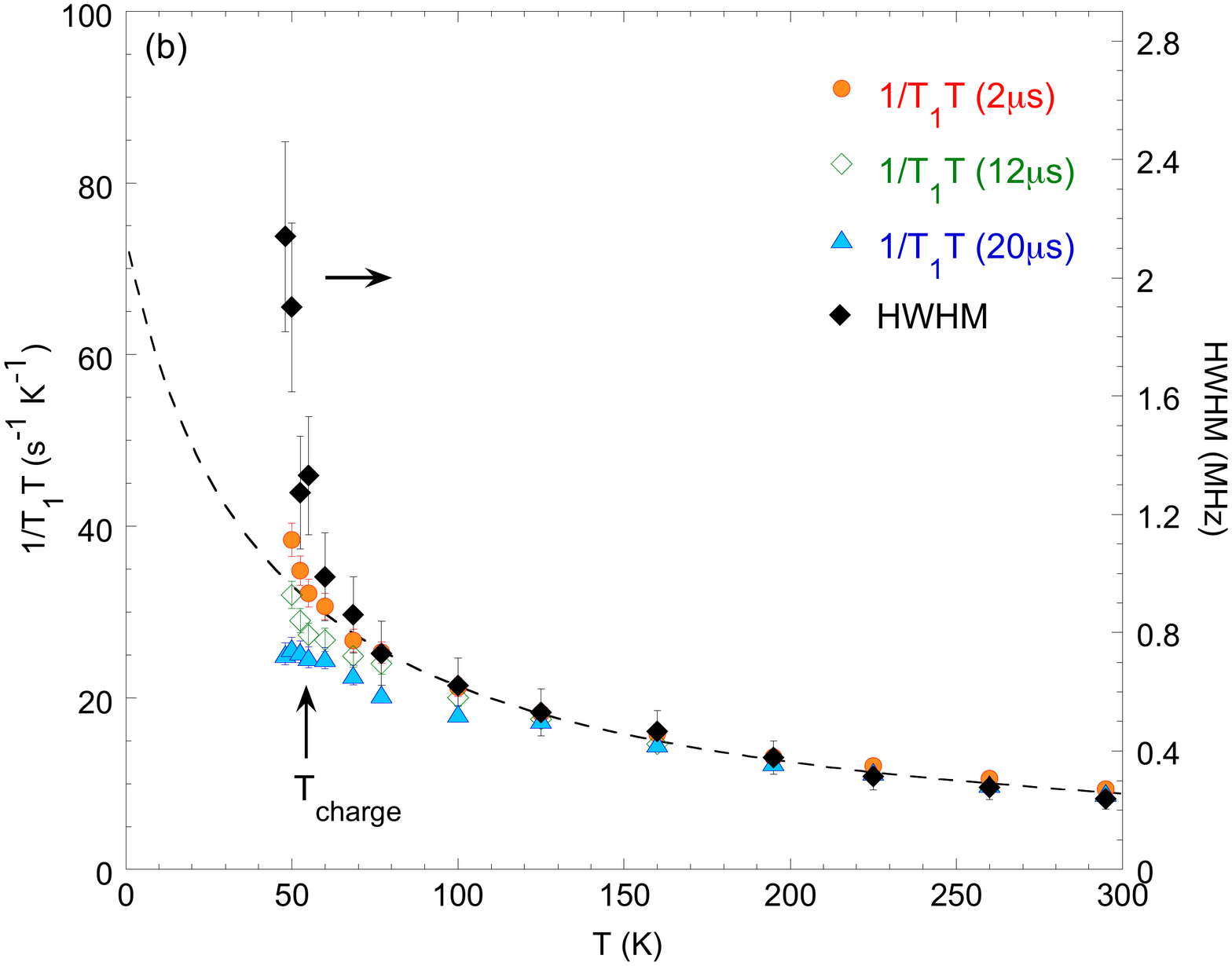}
		\caption{(a) $1/T_{1}$ measured for $^{63}$Cu at the peak of the c-axis lineshape using three different values of the pulse separation time $\tau = 2$, 12, and 20~$\mu$s.  Downward open triangles are the results measured with $\tau = 20$~$\mu$s for a uni-axially aligned powder sample \cite{ImaiJPSJ1990}.  Also shown using grey open bullets is the center of gravity of the distributed $^{139}1/T_{1}$ at $^{139}$La sites of the same single crystal, deduced from inverse Laplace transform of the nuclear spin recovery curve \cite{SingerPRB2019}, whereas the dashed curves below 100~K represent the top or bottom 10\% value of the distributed $^{139}1/T_1$.  (b) $1/T_{1}T$, presented with the same symbols as in the panel (a).  Also shown using the right axis is the HWHM of the c-axis lineshapes in Fig.\ref{fig:lineshape}(b).   The dashed curve represents the Curie-Weiss fit at higher temperatures with Weiss temperature $-40$~K.  Extrapolation of the fit below $\sim 80$~K underestimates the HWHM by $\sim 40$\% at $T_{\text{charge}}$.  }
		\label{fig:T1}
	\end{center}
\end{figure}

In Fig.\ref{fig:T1}(b), we summarize the temperature dependence of the half width at the half maximum (HWHM) of the c-axis lineshapes shown in Fig.\ref{fig:lineshape}(b).   For the $B_{\text{ext}}~||~$~c-axis geometry, the second order nuclear quadrupole effect vanishes \cite{PenningtonPRB1989tensor}, and the temperature dependence of the linewidth is set almost entirely by magnetic effects \cite{FootnotenuQwidth}.  On the other hand, since the lower frequency side of the broadened NMR lineshape nominally has negative frequency shifts, a large distribution of the chemical shift cannot account for the observed broadening, either.  Therefore, the spin degrees of freedom must be playing the key role in the line broadening, but the exact mechanism of the broadening has long been an enigma. 

The dashed curve overlaid on the HWHM data points above $\sim 80$~K are the best empirical Curie-Weiss fit.  The HWHM begins to grow more quickly below $\lesssim 80$~K.   Our single crystal result is consistent with an earlier aligned powder result \cite{GotoPhysicaB1994}.  This linewidth anomaly is accompanied by an analogous deviation from the Curie-Weiss growth of $1/T_{1}T$ at $^{63}$Cu sites \cite{TouLBCOT1}, signaling that strong enhancement of antiferromagnetic spin correlations is playing a role in HWHM as well.  Interestingly, charge order sets in for a minor volume of La$_{1.885}$Sr$_{0.115}$CuO$_{4}$ also at $\sim 80$~K and the aforementioned wing-like $^{63}$Cu NMR signal emerges \cite{ImaiPRB2017}, but it may be a coincidence.   

We recently showed based on the ILTT$_{1}$ analysis of the $^{139}$La nuclear spin-lattice relaxation curve that the electric field gradient (EFG) at the $^{139}$La sites has slowly fluctuating components ($\sim$~MHz) below $\sim 80$~K \cite{SingerPRB2019}.  The ILT cannot distinguish the origin of the slow dynamics between the lattice and/or charge degrees of freedom.  Regardless of the origin, these NMR results indicate that spin correlations begin to grow more steeply when fluctuations of the lattice and/or charge degrees of freedom slow down.  It is also interesting to note that recent X-ray scattering data showed the dynamic short range charge order above $T_{\text{charge}}$ \cite{MiaoPRX2019}.  All pieces put together, the onset of charge order in La$_{1.875}$Ba$_{0.125}$CuO$_{4}$ seems to be {\it suppressed} to $\sim 54$~K, until the LTO to LTT structural phase transition suddenly takes place. 

\subsection{$^{63}$Cu spin-lattice relaxation rate $^{63}1/T_{1}$}
We measured $^{63}1/T_{1}$ at the center of the $B_{\text{ext}}~||~c$~axis peak in Fig.\ \ref{fig:lineshape}(b) using the standard inversion recovery method by applying a 180 degree pulse prior to the spin echo sequence.   The goodness of the fit of the nuclear spin recovery curve $M(t)$ with the standard formula for the central transition was similar to the case of La$_{1.885}$Sr$_{0.115}$CuO$_{4}$ \cite{ImaiPRB2017}, and stretching was not necessary for our purpose even below $T_{\text{charge}}$ owing to modest distributions, as shown in Appendix.  This is simply because the signals arising from $^{63}$Cu sites with very fast $^{63}1/T_1$ are wiped out below $T_{\text{charge}}$ even for $\tau = 2$~$\mu$s (see Fig.\ref{fig:T2} below), and hence hardly contribute to the $M(t)$ results.   In Fig.\ref{fig:T1}(a), we summarize the temperature dependence of $^{63}1/T_{1}$ observed for three different values of the pulse separation time $\tau = 2$, 12, and 20~$\mu s$ between the 90 and 180 degree pulses, and compare the results with $^{139}1/T_1$ observed at the $^{139}$La sites \cite{SingerPRB2019}.    

The gradual decrease of $^{63}1/T_{1}$ with temperature observed for $\tau = 20$~$\mu$s down to 48~K is typical for high $T_c$ cuprates \cite{ImaiJPSJ1988-2}.  The extremely broad, small NMR signal from a small single crystal made accurate measurements of $^{63}1/T_1$ difficult below 48~K.  For comparison, we show the $^{63}1/T_1$ results measured with $\tau = 20$~$\mu$s for an aligned powder sample \cite{ImaiJPSJ1990}.  The signal intensity was large and manageable even below 48~K for the large amount ($\sim 300$~mg) of aligned powder, and the decreasing trend of $^{63}1/T_1$ continues below 48~K.  These results for $\tau = 20$~$\mu$s show no anomaly through $T_{\text{charge}}$, and indicate that some parts of CuO$_2$ planes remain unaffected by charge order and the resulting enhancement of spin fluctuations even deep into the charge ordered state below $T_{\text{charge}}$.  That is why we initially overlooked the lurking charge order in 1990 \cite{ImaiJPSJ1990}.

The $^{63}1/T_{1}$ results for $\tau = 2$~$\mu$s are consistently larger by $\sim 6$\% than those for $\tau = 20$~$\mu$s down to $\sim 80$~K.  This is merely because the quenched disorder caused by Ba$^{2+}$ substitution into the La$^{3+}$ induces a nanoscale inhomogeneity in the local hole concentration of the CuO$_2$ planes \cite{SingerPRL2002, SingerPRB2005}, as represented schematically by different shades in Fig.\ref{fig:domains}(a).  $^{63}1/T_{1}$ is generally smaller for larger values of the hole concentration $x$ \cite{ImaiPRL1993_1, SingerPRL2002}, but the $^{63}$Cu NMR peak frequency for the $B_{\text{ext}}~||~c$~axis geometry is set entirely by the chemical shift that is independent of $x$.  Moreover, the $^{63}$Cu B-sites located at the nearest neighbor of La$^{3+}$ sites are superposed in this field geometry \cite{Yoshimura1992}, and their $^{63}1/T_{1}$ is somewhat slower  than those at the main $^{63}$Cu A-sites \cite{SingerPRL2002}.  Accordingly, a mild distribution of $^{63}1/T_1$ is always present.  As shown below in Fig.\ref{fig:T2}(a), the transverse relaxation does not reduce the spin echo intensity significantly for $\tau = 2$~$\mu$s above $T_{\text{charge}}$, and hence nearly 100\% of the $^{63}$Cu nuclear spins contribute to the observed value of $^{63}1/T_{1}$.

$^{63}1/T_{1}$ for $\tau = 2$~$\mu$s levels off towards $T_{\text{charge}}$, and shows qualitatively different behavior from the $\tau = 20$~$\mu$s results.  This is consistent with the increased distribution of $^{139}1/T_{1}$ observed at $^{139}$La sites in the same temperature range \cite{SingerPRB2019, BaekLaT1PRB2015}.  $^{63}1/T_{1}$ measured with $\tau = 2$ and $12$~$\mu$s begins to increase precisely below $T_{\text{charge}}$, in agreement with the earlier NQR report by Tou et al. \cite{TouLBCOT1}.  Pelc et al. observed greater values of $^{63}1/T_{1}$ below $T_{\text{charge}}$ with NQR than Tou et al., because they used $\tau = 2$~$\mu$s and captured more nuclear spins with faster relaxation rates \cite{PelcPRB2017}.  Our $^{63}1/T_{1}$ results for $\tau = 2$~$\mu$s is slower below $T_{\text{charge}}$ than Pelc et al.'s, probably because the fastest nuclear spins are pushed aside to the tail sections of the magnetically broadened NMR lineshape below $T_{\text{charge}}$ due to locally stronger spin correlations.  

We present $^{63}1/T_{1}T$ in Fig.\ref{fig:T1}(b) in comparison to the HWHM.  $^{63}1/T_{1}T$ probes the wave vector ${\bf q}$ integral of Im$\chi({\bf q}, \omega_{n})$.  $^{63}1/T_{1}T$ obeys the Curie-Weiss behavior analogous to that observed for the HWHM.  The $^{63}1/T_{1}T$ results for $\tau = 2$~$\mu$s as well as the HWHM begin to deviate from the Curie-Weiss behavior somewhat above $T_{\text{charge}}$, in agreement with the earlier report by Tou et al. as noted above \cite{TouLBCOT1}.

The enhancement of $^{63}1/T_{1}$ observed at $^{63}$Cu sites below $T_{\text{charge}}$ is only modest, compared with the steep divergent behavior observed at $^{139}$La sites for the same crystal (grey bullets) \cite{DifferentT1atLa}.  This apparent discrepancy arises from the fact that $^{139}1/T_{1}$ plotted for the $^{139}$La sites represents the spatially averaged (center of gravity) value of the widely distributed $^{139}1/T_{1}$ below $T_{\text{charge}}$.  In contrast, $^{63}1/T_{1}$ at $^{63}$Cu sites reflects {\it only the nuclear spins that are still observable below $T_{\text{charge}}$ owing to their slower NMR relaxation rates}.  For example, the observable $^{63}$Cu NMR signal intensity at 48~K is only about a half of the total intensity even for $\tau =2$~$\mu$s, because a majority of $^{63}$Cu NMR signals is already suppressed by their extremely fast transverse relaxation rates (see the data points at $2\tau = 4$~$\mu$s in Fig.\ref{fig:T2}(b) below).   

To underscore this point, we used dashed lines in Fig.\ref{fig:T1}(a) to mark the top and bottom 10\% values of the distributed $^{139}1/T_{1}$ at $^{139}$La sites estimated from the ILTT$_1$ analysis \cite{SingerPRB2019}.  The comparison indicates that $1/T_{1}$ measured at the observable $^{63}$Cu sites below $T_{\text{charge}}$, especially for the longer values of $\tau$, reflects only the bottom end of the spatial distribution in spin fluctuations.  

\begin{figure}
	\begin{center}
		\includegraphics[width=3in]{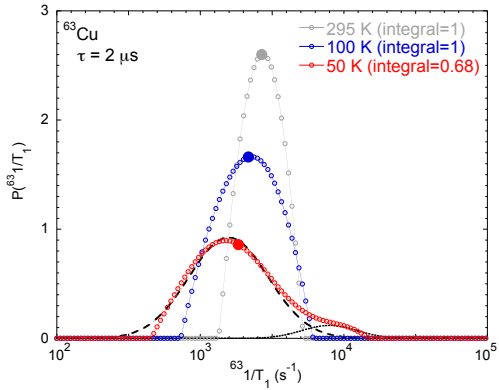}
		\caption{The density distribution function $P(^{63}1/T_{1})$ of the distributed values of $^{63}1/T_{1}$ at $^{63}$Cu sites, deduced by ILT from $M(t)$ measured with $\tau = 2$~$\mu$s shown in Appendix.  The integrated area (in a log scale) is normalized to 1 at 295~K and 100~K (so that the total probability is 1), whereas the area underneath the 50~K result is set to 0.68 in proportion to the suppressed signal intensity observed at 50~K with $\tau = 2$~$\mu$s (see Fig.\ref{fig:T2}(a) below).  The filled bullets mark the center of gravity of $P(^{63}1/T_{1})$.   Dashed and dotted lines represent deconvolution of $P(^{63}1/T_{1})$ curve at 50~K with two Gaussians associated with the canonical domains with slower $^{63}1/T_{1}$ and charge ordered domains with enhanced $^{63}1/T_{1}$, respectively.   The integral of the dashed curve is $0.63$, implying that the volume fraction $F_{\text{CA}}$ of the canonical domains at 50~K($<T_{\text{charge}}$) is still finite at $\sim$63\% even though charge ordering is already under way.  The integral of the dotted curve, 0.05, is much smaller than 1 - 0.63 = 0.37, and severely underestimates the fraction of charge ordered domains $F_{\text{CO}}$; this is simply because their contribution to the $M(t)$ curve is suppressed by fast $T_2$.
		}
		\label{fig:CuILT}
	\end{center}
\end{figure}

These findings can be corroborated by the density distribution function $P(^{63}1/T_{1})$ at $^{63}$Cu sites deduced by ILT.  In Fig.\ref{fig:CuILT}, we summarize $P(^{63}1/T_{1})$ obtained from the $M(t)$ curves measured with $\tau = 2$~$\mu$s in Fig.~9(a).  The integrated area underneath the $P(^{63}1/T_{1})$ curve for 50~K is set to 0.63 to reflect the suppressed signal intensity observed with $\tau = 2$~$\mu$s at 50~K, as shown in Fig.~6(a) in the next section.     In general, the stretched fit value of $1/T_1$ is merely a crude approximation of the center of gravity of the distribution $P(1/T_{1})$ \cite{SingerPRB2019}.  In fact, we found that the center of gravity $^{63}1/T_{1}=1840$~s$^{-1}$ of $P(^{63}1/T_{1})$ at 50~K is close to the stretched fit value (1920~s$^{-1}$) plotted in Fig.\ref{fig:T1}.  It is important to notice, however, that $^{63}1/T_1$ at 50~K has both the slower and faster components than 100~K.   The fast components reach as large as $\sim10^{4}$~s$^{-1}$, in agreement with the expectations from the divergent growth of $^{139}1/T_{1}$ below $T_{\text{charge}}$.  On the other hand, the slow components extends to below $10^{3}$~s$^{-1}$, again in agreement with the expectations from $^{63}1/T_1$ measured with $\tau = 20$~$\mu$s.  

\subsection{Spin echo decay}
\begin{figure}
	\begin{center}
		\includegraphics[width=3in]{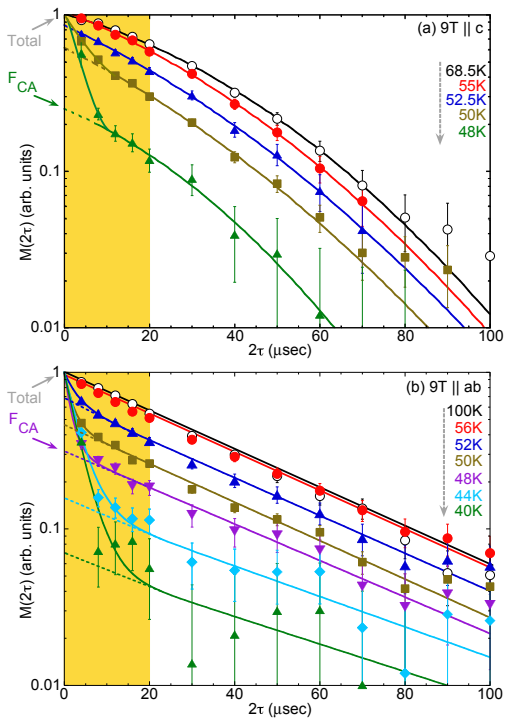}
		\caption{Representative spin echo decay curves $M(2\tau)$ observed at the peak of the lineshapes in Fig.\ref{fig:lineshape} for a magnetic field applied along the (a) $B_{\text{ext}}~||$~c-axis and (b)$B_{\text{ext}}~||$~ab plane.  (We confirmed that the results for $^{63}$Cu(B) sites with $B_{\text{ext}}~||$~ab plane are very similar.)  The overall intensity at different temperatures is normalized by the integrated intensity of the lineshape measured with $\tau = 12$~$\mu$s and the Boltzmann factor.  Solid curves are the best fit above $2\tau = 20$~$\mu$s with (a) Gaussian-Lorentzian and (b) Lorentzian (i.e. exponential) function.  Below $2\tau = 20$~$\mu$s, as a guide for eyes, we smoothly extrapolated the solid lines to $M(2\tau = 0) = 1$.  Dashed curves in both panels represent the extrapolation of the fit in the long time regime to $2\tau = 0$, corresponding to the dashed lines in Fig.\ref{fig:response}(c).  The intercept of the dashed curves with the vertical axis at $2\tau = 0$ yields the volume fraction of the canonical domains, $F_{\text{CA}}$.}
		\label{fig:T2}
	\end{center}
\end{figure}

In Fig.\ref{fig:T2}, we summarize the representative spin echo decay curves $M(2 \tau)$ observed at the peak of the NMR lineshapes in Fig.\ref{fig:lineshape}.  We normalized the magnitude of $M(2 \tau)$ with the integral of the lineshape in Fig.\ref{fig:lineshape} as well as the Boltzmann factor.  Thus $M(2 \tau)$ in the limit of $2 \tau = 0$ represents the temperature independent {\it total intensity} as represented schematically by the grey horizontal line in Fig.\ref{fig:response}(d).  For clarity, we normalize the total intensity to $M(2\tau=0) = 1$.

The $B_{\text{ext}}~||$~c results in Fig.\ref{fig:T2}(a) show the typical Gaussian-Lorentzian decay form for the long time regime above $2\tau \sim 20$~$\mu$s both above and below $T_{\text{charge}}$; the indirect nuclear spin-spin coupling causes the Gaussian curvature associated with the real part Re$\chi({\bf q}, \omega)$ of the dynamical electron spin susceptibility of Cu \cite{PenningtonPRB1989, PenningtonPRL1991}.  For the $B_{\text{ext}}~||$~ab geometry shown in Fig.\ref{fig:T2}(b), this Gaussian contribution is motionally narrowed to Lorentzian \cite{PenningtonPRB1989}.  In addition, Redfield's $T_{1}$ process associated with the imaginary part Im$\chi({\bf q}, \omega)$ of the dynamical electron spin susceptibility of Cu leads to the exponential, Lorentzian process in both field geometries.  Since Re$\chi({\bf q}, \omega)$ and Im$\chi({\bf q}, \omega)$ are related to each other, $1/T_{1}T$ and $1/T_2$ in cuprates generally exhibit analogous temperature dependences  \cite{ItohJPSJ1992, ImaiPRB1993, ImaiPRL1993_2}, unless the pseudo-gap strongly suppresses only $1/T_{1}T$ \cite{ItohJPSJ1992}.  

As explained above using Fig.\ref{fig:response}(c), (d), (g), and (h), $M(2\tau = 0)$ in the limit of $2\tau = 0$ is proportional to the number of nuclear spins in our sample, and is a conserved quantity.  Above $T_{\text{charge}}$, the extrapolation of $M(2 \tau)$ curves to $2\tau = 0$ based on the Gaussian-Lorentzian and Lorentzian fit for the $B_{\text{ext}}~||$~c and $B_{\text{ext}}~||$~ab geometry, respectively, is consistent with such expectations.  

Notice, however, that the situation completely changes once charge order sets in at $T_{\text{charge}}$.  $M(2 \tau)$ begins to exhibit a very fast initial decay from $2\tau = 0$ up to $2\tau \sim 15$~$\mu$s for both field geometries.  For example, as shown by the green solid line in Fig.\ref{fig:T2}(a), the $M(2 \tau)$ measured at 48~K in $B_{\text{ext}}~||$~c decays quickly from $M(2 \tau = 0) = 1$ to $M(2 \tau = 10~\mu s) \sim 0.2$ with positive curvature, followed by much slower Gaussian-Lorentzian decay above $2 \tau = 15~\mu$s with the relaxation times comparable to those observed above $T_{\text{charge}}$.  The crossover of $M(2 \tau)$ from the short to long time regime is observed only below $T_{\text{charge}}$, and indicate emergence of the distributed fast transverse $T_{2}$ relaxation processes in the charge ordered state.  This corresponds to the analogous crossover depicted in Fig.\ref{fig:response}(c).  That is, {\it the CuO$_2$ planes develop strong inhomogeneous spin correlations as soon as charge order sets in}.

The $M(2 \tau)$ results in Fig.\ref{fig:T2} are also consistent with the reduction of the signal intensity below $T_{\text{charge}}$ observed at a fixed $2\tau = 24$~$\mu$s in the inset of Fig.\ref{fig:lineshape}.  That is, the signal loss reflects the emergence of glassy spin state in the charge ordered domains, where the NMR relaxation rates $1/T_{1}$ and $1/T_{2}$ become divergently large, resulting in the initial quick decay in the spin echo intensity $M(2 \tau)$.  

We note that our original publications two decades ago probed $M(2\tau)$ only in the long time regime above $2\tau \simeq 20$~$\mu$s \cite{HuntPRL1999} because the spectrometer dead time prevented us from accessing the short time regime in Fig.\ref{fig:T2} and Fig.\ref{fig:response}(c).  Our original $M(2\tau)$ data barely missed the crossover from the short to long time regime \cite{HuntPRL1999}.  On the other hand, a recent work by Pelc et al. presented $M(2\tau)$ data only in the short time regime, and did not demonstrate the crossover to the long time regime, either \cite{PelcPRB2017} (see their Fig.\ 3(a)).  Pelc et al.'s limited data set might inadvertently leave unsuspecting readers with a false impression that CuO$_2$ planes in charge ordered La$_{1.875}$Ba$_{0.125}$CuO$_{4}$ is spatially homogeneous, and exhibit uniformly fast transverse relaxation, similar to the case of homogeneous antiferromagnet shown in Fig.\ref{fig:response}(g).  But our new data presented in Fig.\ref{fig:T2} firmly establish that is not the case.  

We also emphasize that what matters in understanding the inhomogeneous glassy state induced by charge order based on $^{63}$Cu NMR intensity is the extrapolation of the {\it slowly decaying part} of $M(2\tau)$ observed in the long time regime {\it above} $2\tau = 20$~$\mu$s to $2\tau = 0$~$\mu$s, as explained in section 2 and shown with the dashed curves in both Fig.\ref{fig:response}(c) and Fig.\ref{fig:T2}.  Instead, Pelc et al.  examined the extrapolation of the {\it fast} decaying part of their $M(2\tau)$ data in the short time regime to $2\tau = 0$ (solid curves extrapolated to $2\tau = 0$ in their Fig.\ 3(a)), only to confirm that the {\it total intensity} arising from both the charge ordered and canonical domains is conserved.  Their finding below $T_{\text{charge}}$ summarized in the inset of their Fig.\ 2 corresponds to the trivial conservation law of the {\it total intensity} $F_{\text{CO}}+F_{\text{CA}} = 1$, represented schematically by the gray horizontal line in Fig.\ref{fig:response}(d).  It does not provide any useful insight into the nature of the glassy, charge ordered state.

\subsection{$^{63}$Cu NMR signal intensity wipeout and estimation of $F_{CA}$ based on ILT}

\begin{figure}[b]
	\begin{center}
		\includegraphics[width=3in]{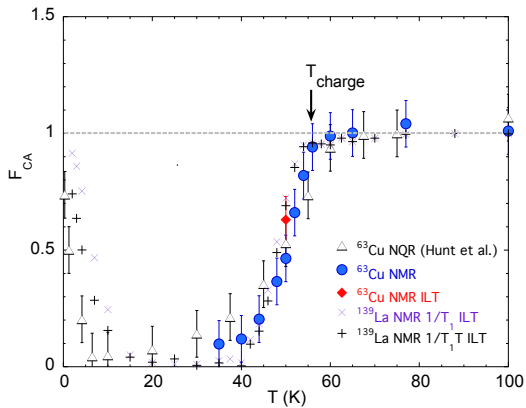}
		\caption{The volume fraction $F_{\text{CA}}$ of the canonical domains below $T_{\text{charge}}$ estimated by three different methods. (bullets): the estimation from $^{63}$Cu NMR signal intensity of the canonically behaving sites with slow relaxation rates, based on the extrapolation of $M(2\tau)$ curves from $2\tau=24$~$\mu$s and greater.  For comparison, we also reproduce the powder NQR intensity (black triangles, adopted from \cite{HuntPRB2001}).  The increase in the NQR intensity below $\sim 15$~K is due to the freezing of the fluctuations of the hyperfine magnetic fields from Cu electron spins; we multiplied a factor of 1.63 \cite{HuntPRB2001} for the Zeeman perturbed NQR results below 15~K to account for the missing contribution below 25~MHz \cite{HuntPRB2001}.  (red diamond): $F_{\text{CA}}$ at 50~K estimated from the $^{63}$Cu ILT result in Fig.\ref{fig:CuILT}.  (Purple $\times$): $F_{\text{CA}}$ estimated from the ILTT$_1$ analysis of the $^{139}$La ILT results in Fig.\ref{fig:T1ILT}.  Also shown with + symbols is $F_{\text{CA}}$ estimated from $P(^{139}1/T_{1}T)$.}
		\label{fig:wipeout}
	\end{center}
\end{figure}

Finally but not the least, let us return to the issue of the integrated intensity of the $^{63}$Cu NMR lineshapes in the inset of Fig.\ref{fig:lineshape}.  To eliminate the minor effects of the transverse relaxation on the bare integrated intensity measured at a fixed $\tau = 12$~$\mu$s, we extrapolated the $M(2\tau)$ curves to $2\tau = 0$ as shown by the dashed lines in Fig.\ref{fig:T2}(b), and estimated the volume fraction $F_{\text{CA}}$ of the canonical domains.  We summarize the $B~||~ab$-axis results in Fig.\ref{fig:wipeout} using bullets \cite{c-axis}.   For comparison, we also plot our original signal intensity wipeout data measured with NQR for a $^{63}$Cu isotope enriched powder sample (open triangles) \cite{HuntPRB2001}.  The agreement between the new NMR and older NQR results is satisfactory, in view of the greater uncertainties in the latter arising from the  extra Gaussian $T_{\text{2G}}$ term in the spin echo decay.

As explained in detail in section 2 using Fig.\ref{fig:response}(d), the temperature dependence of the $^{63}$Cu NMR signal intensity in Fig.\ref{fig:wipeout} indicates that charge order does not set in homogeneously in the CuO$_2$ planes.  The finite value of $F_{\text{CA}}$ below $T_{\text{charge}}$ implies that a significant fraction of the volume is hardly affected by charge order even below $T_{\text{charge}}$, and exhibits the canonical behavior expected for CuO$_2$ planes that seem destined to undergo superconducting transition at $T_{\text{c}} \simeq 30$~K.  But the residual volume fraction of such canonically behaving CuO$_2$ planes almost vanishes by $\sim 35$~K, where the charge correlation length saturates at $\sim$20~nm \cite{TranquadaPRB2008}.  In addition, spin stripe order sets in at $\sim 35$~K at the time scale of $\mu$SR experiments \cite{Luke1991, Nachumi} and the volume-averaged value of the distributed $^{139}1/T_1$ is peaked at $^{139}$La sites \cite{SingerPRB2019, BaekLaT1PRB2015}.  In contrast, the volume fraction of the canonical domains exceeds 40~\% at its $T_{\text{c}}=30$~K in La$_{1.885}$Sr$_{0.115}$CuO$_{4}$ \cite{ImaiPRB2017, ArsenaultPRB2019, SingerPRB2019}.

\begin{figure}[t]
	\begin{center}
		\includegraphics[width=3in]{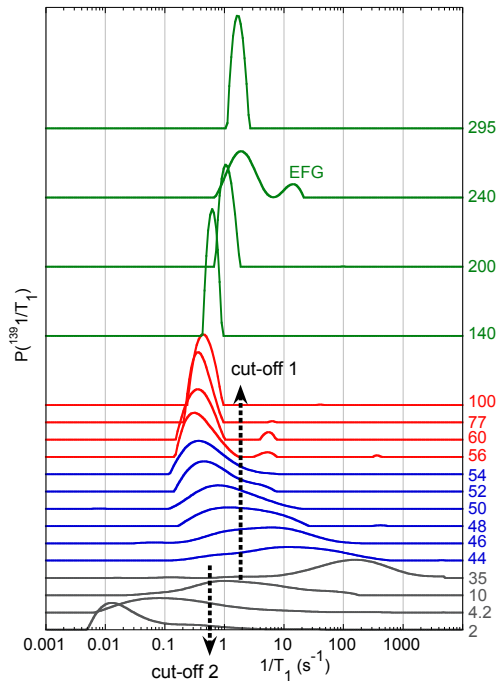}
		\caption{The probability density $P(^{139}1/T_{1})$ (i.e. the histogram of the distributed $^{139}1/T_{1}$), generated from inverse Laplace transform (ILT) of $^{139}$La nuclear spin-lattice relaxation curve (adopted from \cite{SingerPRB2019}).  The black dashed vertical arrows represent the cut-off 1 and 2 set at $^{139}1/T_{1} = 2$~s$^{-1}$ and at $^{139}1/T_{1} = 0.6$~s$^{-1}$, respectively, used for estimating $F_{\text{CA}}$ above and below 30~K.  A small split-off peak observed at 240~K (marked by EFG), accompanied by the broadening of the main peak, is caused by slow fluctuating electric field gradient due to the structural phase transition.  Analogous features manifest slightly above $T_{\text{charge}}$ as well.	}
		\label{fig:T1ILT}
	\end{center}
\end{figure}

We can achieve more quantitative understanding of the $^{63}$Cu NMR intensity anomaly and its relation with the unconventional nature of charge order with the aid of the ILTT$_1$ analysis of the $^{139}$La nuclear spin recovery curve \cite{SingerPRB2019}.  For convenience, we reproduce the key results of the probability density distribution function, $P(^{139}1/T_{1})$ of $^{139}1/T_{1}$ in Fig.\ref{fig:T1ILT}.  The main peak of $P(^{139}1/T_{1})$ has finite values only below $^{139}1/T_{1} \simeq 1$~s$^{-1}$ from 100~K down to 60~K.  In other words, the upper bound of the distributed values of $^{139}1/T_{1}$ is 1~s$^{-1}$.  Notice, however, that the main peak gradually broadens below 77~K, accompanied by a small split-off peak centered around $^{139}1/T_{1} \simeq 5$~s$^{-1}$.  Analogous anomalies of $P(^{139}1/T_{1})$ are observed also around 240~K near the high temperature tetragonal to low temperature orthorhombic structural phase transition \cite{SingerPRB2019}.  Since the charge order transition is accompanied by a first order structural transition from low temperature orthorhombic to low temperature tetragonal phase \cite{Fujita2004}, we can  attribute these anomalies slightly above $T_{\text{charge}}$ to the contributions of fluctuating electric field gradient precursor to the structural phase transition and/or fluctuating charges\cite{SingerPRB2019}.

As temperature is lowered through $T_{\text{charge}}$, $P(^{139}1/T_{1})$ gradually transfers spectral weight to larger values of $^{139}1/T_{1}$ while broadening asymmetrically.  This corresponds to the fact that a sharp divergent behavior sets in precisely at $T_{\text{charge}}$ for $^{139}1/T_{1}$ estimated from the stretched fit, which tends to be close to the center of gravity of the distributed $^{139}1/T_{1}$ \cite{SingerPRB2019}.  We emphasize that a half of the spectral weight of $P(^{139}1/T_{1})$ still remains below 1~s$^{-1}$ even at 50~K. This implies that the corresponding sample volume is still unaffected by charge order, and $^{139}1/T_{1}$ is as slow as at 77 K.  This is consistent with our findings for $^{63}1/T_{1}$ at $^{63}$Cu sites in Fig.\ref{fig:T1} and \ref{fig:CuILT} . $^{63}$Cu nuclear spins that are still easily observable below $T_{\text{charge}}$ owing to slow NMR relaxation rates are located in the same domains as these $^{139}$La sites with slower relaxation rates. 

In view of the fact that the ILT curve $P(^{63}1/T_{1})$ observed at 50~K in Fig.~\ref{fig:CuILT} has a well-defined peak associated with the canonical domains with slower $^{63}1/T_{1}$, perhaps it may be somewhat surprising to find that $P(^{139}1/T_{1})$, which should also encompass the canonical component centered around $^{139}1/T_{1} \simeq 0.5$~s$^{-1}$, is broader and increasingly featureless below $T_{\text{charge}}$.   But this is simply because  the transverse relaxation does not suppress the faster components of $P(^{139}1/T_{1})$ arising from charge ordered domains.  In this context, we recall that the charge correlation length in the charge ordered state is known to be as short as several nm immediately below $T_{\text{charge}}$, and the spin correlation length cannot exceed it.  This means that the extent of enhancement of $1/T_1$ in each charge ordered domain is set by the domain size.  The highly disordered nature of the charge ordered state with varying domain sizes naturally explains the very broad distribution of $^{139}1/T_1$ below $T_{\text{charge}}$, ranging from the small canonical value to the upper bound set by the largest charge ordered domains.  It is also worth noting that $P(^{139}1/T_{1})$ curve exhibits somewhat more distinct features in La$_{1.885}$Sr$_{0.115}$CuO$_{4}$ for the canonical and charge ordered domains\cite{ArsenaultPRB2019}.  That is probably because the canonical domains are more robust below $T_{charge}$ in a wider temperature range in La$_{1.885}$Sr$_{0.115}$CuO$_{4}$, and in agreement with the fact that $T_{c}$ is as high as $\sim30$~K. 

The featureless, continuous distribution of $P(^{139}1/T_{1})$ makes it difficult to de-convolute $P(^{139}1/T_{1})$ and estimate $F_{\text{CA}}$ from $P(^{139}1/T_{1})$.  We therefore introduce a cut off in Fig.\ref{fig:T1ILT} at $2$~s$^{-1}$, at the upper end of the distributed values of $^{139}1/T_{1}$ observed at 56~K, as represented by the upward vertical arrow in Fig.\ref{fig:T1ILT}.  Then we can estimate the $F_{\text{CA}}$ of the canonical $^{63}$Cu nuclear spins as the integrated area of  $P(^{139}1/T_{1})$ below the cut-off.  We summarize the temperature dependence of thus estimated $F_{\text{CA}}$ in Fig.\ref{fig:wipeout} using $\times$ symbols, in comparison to $F_{\text{CA}}$ estimated from the $^{63}$Cu NMR intensity.  Despite the simplicity of this analysis, the estimation based entirely on the $^{139}$La NMR results reproduces the $^{63}$Cu NMR signal intensity anomaly very well.

We can also test the consistency of $F_{\text{CA}}$ with the $^{63}$Cu ILT result of $P(^{63}1/T_{1})$ at 50~K.  From the integral of the light dashed curve  in Fig.\ref{fig:CuILT} arising from the canonical contribution with slower relaxation rate, we estimate $F_{\text{CA}} \sim 0.63$ at 50~K.  We plot the result in Fig.\ref{fig:wipeout} with a diamond, in comparison to $F_{\text{CA}}$ estimated from two other methods, the extrapolation of $M(2\tau)$ (bullets) and cut-offs introduced for $P(^{139}1/T_{1})$ ($\times$).  Despite the completely different methodologies between the three approaches, agreement is good. 

Turning our attention to the low temperature side below 30~K,  Zeeman perturbed NQR signal is known to reemerge when the hyperfine magnetic field from frozen Cu electron spins become static at the NMR measurement time scale below $\sim 15$~K \cite{TouLBCOT2, HuntPRB2001}. In general, $1/T_1$ is proportional to the dynamical spin susceptibility Im$\chi(\omega)$ multiplied by temperature $T$, and hence the cut-off of $^{139}1/T_{1}=2$~s$^{-1}$ for the same magnitude of Im$\chi(\omega)$ needs to be scaled down to $^{139}1/T_{1}=0.6$~s$^{-1}$ by the ratio between 15~K and $T_{\text{charge}}$.  This second cut-off is shown with a downward vertical arrow in Fig.\ref{fig:T1ILT}.  We can estimate the fraction of frozen Cu electron spins as the area integral below this cut-off.  The results, also shown in Fig.\ref{fig:wipeout} using $\times$, reproduce the qualitative aspects of the signal intensity recovery observed  by Hunt et al.\cite{HuntPRB2001}.  The agreement can be improved if we estimate $F_{\text{CA}}$ based on $P(^{139}1/T_{1}T)$ (i.e. the distribution of $^{139}1/T_{1}$ divided by $T$) by introducing a single cut-off at $^{139}1/T_{1}T = 0.036$~s$^{-1}$K$^{-1}$ for both above and below 30~K; this cut-off value corresponds to  $^{139}1/T_{1} = 2$~s$^{-1}$ divided by $T_{\text{charge}}$.  We present these estimations using $+$ symbols also in Fig.\ref{fig:wipeout}.

Strictly speaking, the cut-off for dividing the canonical and charge ordered domains at $2$~s$^{-1}$ for $^{139}1/T_{1}$ should be slightly temperature dependent, because $^{139}1/T_{1}$ in canonically superconducting compositions with $x\sim0.15$ decreases slightly below 54~K toward $T_{c}=38$~K~\cite{Kobayashi1989,Yoshimura1992}.  But the observed decrease is weak, and the varying cut-off hardly affects our estimation of $F_{CA}$.  In fact, the cut-off  
$^{139}1/T_{1}T = 0.036$~s$^{-1}$K$^{-1}$ effectively incorporates such temperature dependent shift of the cut-off in $^{139}1/T_{1}$, but the $F_{CA}$ results in Fig.~\ref{fig:wipeout} show no significant changes.

\section{Summary and conclusions}
We reported new comprehensive single crystal $^{63}$Cu NMR results for La$_{1.875}$Ba$_{0.125}$CuO$_{4}$, and compared the results with our recent report on $^{139}$La NMR.  We confirmed the precursors of enhanced growth in spin correlations below $\sim 80$~K based on HWHM and $1/T_{1}T$, in agreement with earlier reports \cite{GotoPhysicaB1994, TouLBCOT1}.  This is the same temperature range, where our recent ILTT$_1$ analysis of $^{139}1/T_1$ at the $^{139}$La NMR sites identified the presence of slow lattice and/or charge fluctuations \cite{SingerPRB2019}.    We demonstrated that the apparently contradictory reports of $1/T_1$ and spin echo decay curves $M(2\tau)$ at the $^{63}$Cu sites near and below $T_{\text{charge}}$, as well as the apparently different behavior between $^{63}$Cu and $^{139}$La sites below $T_{\text{charge}}$, are the consequence of a large spatial distribution in the enhancement of spin fluctuations.

Our findings of the sudden onset of NMR anomalies precisely at $T_{\text{charge}}$ for $M(2\tau)$, $F_{\text{CA}}$, $^{63}1/T_{1}$, and  $^{139}1/T_{1}$ and its asymmetric distribution  are consistent with the earlier inelastic neutron scattering experiments with very small energy transfer, i.e. charge order turns on glassy spin dynamics precisely at $T_{\text{charge}}$ before the static magnetic order sets in at $T_{\text{spin}} \simeq 40$~K, and hence the low frequency Cu spin fluctuations begin to undergo a dramatic enhancement at $T_{\text{charge}}$ \cite{TranquadaPRB59, Fujita2004, TranquadaPRB2008}.  

We revisited our earlier report of the intensity anomaly of $^{63}$Cu NMR below $T_{\text{charge}}$ of La$_{1.875}$Ba$_{0.125}$CuO$_{4}$.  We reproduced our original discovery of the intensity anomaly at $T_{\text{charge}}$ \cite{HuntPRL1999} with much higher precision, by taking advantage of the convenient magnetic field geometry of $B_{\text{ext}}~||~$~ab axis.  We demonstrated once and for all that glassy spin dynamics induced within charge ordered domains begins to suppress the $^{63}$Cu NMR intensity exactly at $T_{\text{charge}}$, where $T_{\text{charge}}$ has already been determined independently by diffraction experiments.  We also explained in detail why the observable fraction $F_{\text{CA}}$ of the $^{63}$Cu NMR signal intensity provides a good measure of the canonical domains that have not been affected significantly by charge order.  Our finding was corroborated by a completely different approach based on the ILTT$_1$ analysis of the distributed $^{63}1/T_1$ and  $^{139}1/T_1$ \cite{SingerPRB2019}.  We recall that we recently achieved the same for La$_{1.885}$Sr$_{0.115}$CuO$_{4}$ based on single crystal $^{63}$Cu NMR and the ILTT$_1$ analysis of $^{139}$La NMR \cite{ImaiPRB2017, ArsenaultPRB2019}.  

We identified a key difference between La$_{1.875}$Ba$_{0.125}$CuO$_{4}$ with $T_{\text{c}}=4$~K and La$_{1.885}$Sr$_{0.115}$CuO$_{4}$ with much higher $T_{\text{c}}=31$~K; charge order enhances spin fluctuations in nearly 100\% volume of the CuO$_2$ planes in the former by $\sim 35$~K, while nearly a half of the volume fraction is still hardly affected when superconductivity sets in at higher $T_{\text{c}}$ in the latter \cite{ImaiPRB2017, ArsenaultPRB2019}.  On the other hand, in view of the fact that the aforementioned anomalous enhancement of spin correlations are commonly observed below $\sim 80$~K for both La$_{1.875}$Ba$_{0.125}$CuO$_{4}$ \cite{TouLBCOT1} and La$_{1.885}$Sr$_{0.115}$CuO$_{4}$ \cite{ImaiJPSJ2018, Mitrovic}, it is not clear why $T_{\text{charge}} \simeq 54$~K is much lower in La$_{1.875}$Ba$_{0.125}$CuO$_{4}$ than $T_{\text{charge}} \simeq 80$~K in La$_{1.885}$Sr$_{0.115}$CuO$_{4}$.  The only signature of charge order for La$_{1.875}$Ba$_{0.125}$CuO$_{4}$ observed to date above $T_{\text{charge}}$ is dynamic in nature \cite{MiaoPRX2019}.  It seems as if charge order in La$_{1.875}$Ba$_{0.125}$CuO$_{4}$ is suppressed from 80~K to 54~K, until the first-order low temperature tetragonal structural transition sets in.  

It may be worthwhile to caution that the $^{63}$Cu NMR intensity anomaly is not always entirely related to static charge order.  In the case of La$_{2-x}$Sr$_{x}$CuO$_{4}$ with \cite{SingerPRB1999} and without \cite{HuntPRL1999} Nd co-doping, we initially attributed the onset of the $^{63}$Cu NQR intensity anomaly to charge order not only for the optimal charge order composition of $x \sim 1/8$, but also for above and below $x \sim 1/8$.  Subsequent X-ray diffraction experiments for La$_{1.6-x}$Na$_{0.4}$Sr$_{x}$CuO$_{4}$ \cite{IchikawaPRL2000} showed that our estimation of $T_{\text{charge}}$ was accurate for $x = 1/8$ and above, but we overestimated $T_{\text{charge}}$ for $x=0.10$ and below.  Our overestimation for $x=0.10$ resulted from the fact that the gradual intensity loss can arise also from localization of doped holes that precedes charge order.  It turned out that the inflection point in the temperature dependence of $F_{\text{CO}}$ at a lower temperature corresponds to $T_{\text{charge}}$ for $x < 1/8$.  We refer readers to Fig.\ 15(a), Fig.\ 18 and related discussions in \cite{HuntPRB2001} for details.\\

\begin{acknowledgments}
We thank J. Wang for helpful discussions.  T. I. is supported by NSERC.   P.M.S. is supported by The Rice University Consortium for Processes in Porous Media.  The work at Tohoku is supported by Grant-in-Aid for Scientific Research (A) (16H02125), Japan.
\end{acknowledgments}

%\bibliography{SingerLBCO_arXiv}% Produces the bibliography via BibTeX.
%\bibliography{ImaiLBCOjpsj_arXiv_tempo}% Produces the bibliography via BibTeX.

\begin{thebibliography}{52}
\bibitem [{Foo()}]{FootnotenuQwidth}%
  \BibitemOpen
  \href@noop {} {\bibinfo {title} {{We confirmed that the linewidth observed
  above and below $T_{charge}$ is proportional to the magntide of
  $B_{\text{ext}}$ for Sr-doped La$_2$CuO$_4$ in [17]. This means that the
  second order quadrupole effect is not the dominant factor of the very broad
  lineshapes in doped La$_2$CuO$_4$}}}\BibitemShut {NoStop}%
\bibitem [{c-a()}]{c-axis}%
  \BibitemOpen
  \href@noop {} {\bibinfo {title} {{The $B~||~c$-axis results are similar (not
  shown), but far less accurate due to the combination of two complications:
  (i) the extra parameter required for fitting the Gaussian contribution
  $T_{\text{2G}}$ in the spin echo decay curve $M(2\tau)$, and (ii) the drastic
  line broadening observed below $T_{\text{charge}}$.}}}\BibitemShut {Stop}%
\end{thebibliography}


\vspace{0.5in}

{\bf Appendix. The recovery curve $M(t)$}\\

\begin{figure}
	\begin{center}
		\includegraphics[width=2.8in]{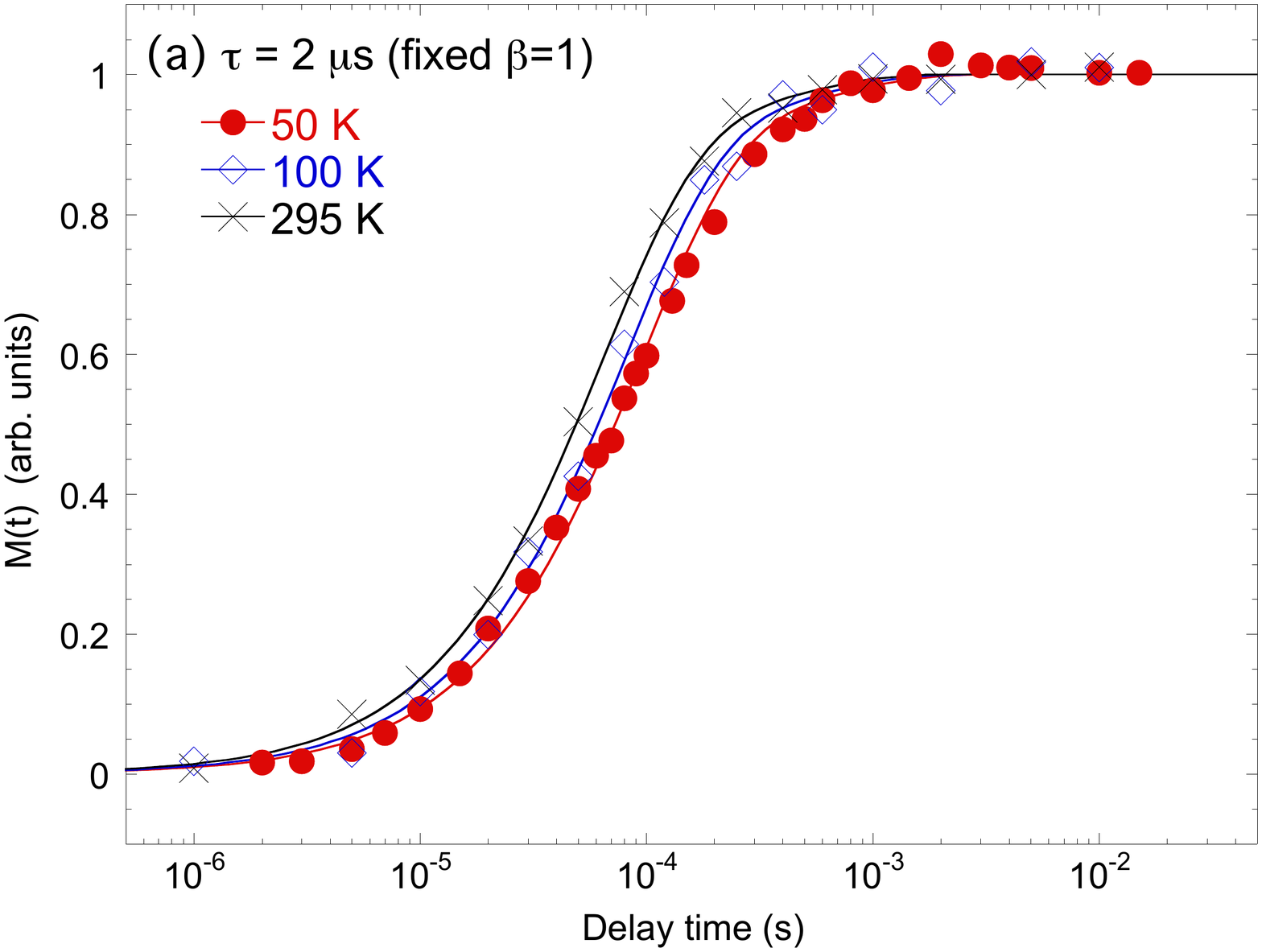}
		\includegraphics[width=2.8in]{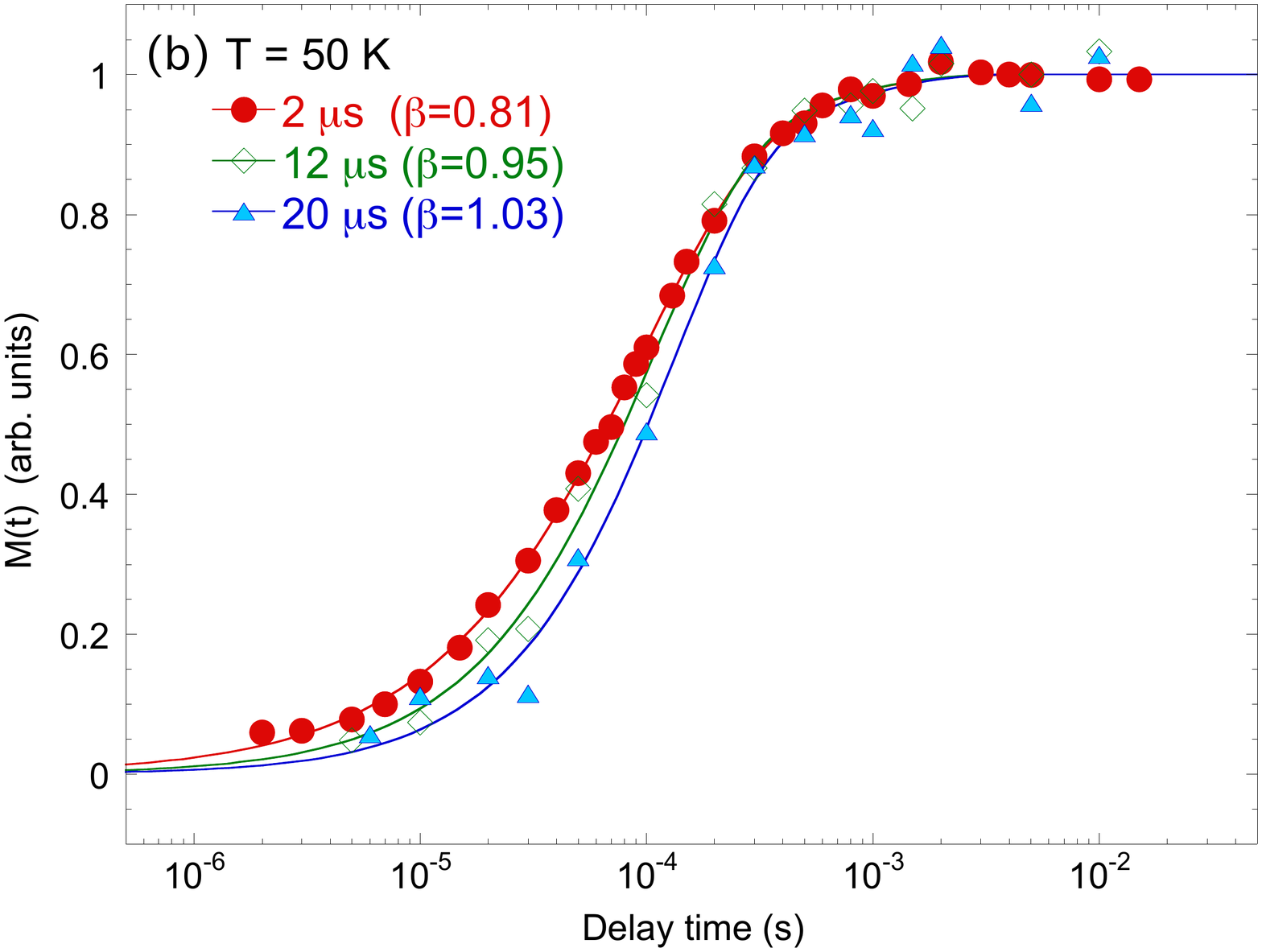}
		\caption{(a) Representative $T_1$ recovery curves $M(t)$ observed for $^{63}$Cu sites with the $B_{\text{ext}}~||$~c geometry at various temperatures, measured with fixed $\tau = 2$~$\mu$s.  The solid curves represent the best fit with fixed $\beta=1$.  (b) Representative $T_1$ recovery curves $M(t)$ at 50~K measured with $\tau = 2$, 12, and 20~$\mu$s.  The solid curves represent the best phenomenological stretched fit, which yielded $\beta = 0.81$, 0.95, and 1.03, respectively.  The signal intensity becomes smaller for longer values of $\tau$ (see Fig.~5(a)), and hence the noise is greater for $\tau = 20$~$\mu$s.    }
		\label{fig:T1M(t)}
	\end{center}
\end{figure}

We measured $^{63}1/T_1$ at the $^{63}$Cu sites using the central transition.  The standard formula for the relaxation recovery $M(t)$ calculated from the coupled rate equations is
\begin{equation}
M(t) = M_{\text{o}}-A[0.9 e^{-(6t/^{63}T_{1})^{\beta}} +0.1 e^{-(t/^{63}T_{1})^{\beta}}],	
\end{equation} 
where $M_{\text{o}}$, $A$, and $^{63}1/T_1$ are the free parameters, and the stretched exponent $\beta$ should be normally set to 1.  We present examples of the normalized $M(t)$ curves measured with fixed $\tau = 2$~$\mu$s in Fig.~9(a), together with the best fit with Eq.(1) for fixed $\beta = 1$.  The standard fit with $\beta = 1$ is sufficiently good even at 50~K below $T_{\text{charge}}$ to illustrate the key aspects summarized in Fig.~4.

Also summarized in Fig.~9(b) is the $\tau$ dependence of $M(t)$ curves observed at 50~K with the phenomenological stretched fit.  Lifting the constraint of $\beta =1$ only marginally improves the fit, and the fitted value of $^{63}1/T_1$ hardly changes.  The overall  relaxation becomes faster for shorter $\tau$, accompanied by greater distribution, as evidenced by the smaller value of $\beta$. But the deviation of $\beta$ from the non-distributed case of 1 is not very significant.  Since the signal intensity becomes very small below $T_{\text{charge}}$ especially for longer $\tau$, we fixed $\beta = 1$ for $^{63}$Cu sites to reduce the number of free fitting parameters in Eq.~(1), and thereby reducing the scattering in the $^{63}1/T_1$ results.

%\nocite{*}


\end{document}